\documentclass[conference]{IEEEtran}
\IEEEoverridecommandlockouts
\usepackage{cite}
\usepackage{amsmath,amssymb,amsfonts}
\usepackage{algorithmic}
\usepackage{graphicx}
\usepackage{textcomp}
\usepackage{xcolor}
\def\BibTeX{{\rm B\kern-.05em{\sc i\kern-.025em b}\kern-.08em
    T\kern-.1667em\lower.7ex\hbox{E}\kern-.125emX}}
    
\begin{document}

\title{Similarities and Learnings from Ancient Literature on Blockchain Consensus and Integrity}

\author{
\IEEEauthorblockN{Ashish Kundu}
\IEEEauthorblockA{
Cisco Research\\
ashkundu@cisco.com}
\and
\IEEEauthorblockN{Arun Ayachitula}
\IEEEauthorblockA{
Kyndryl\\
arun.ayachitula@kyndryl.com}
\and
\IEEEauthorblockN{Nagamani Sistla}
\IEEEauthorblockA{ 
Freelance Researcher\\
sistla.nagamani@gmail.com}
}

%%
%% end of the preamble, start of the body of the document source.


%\renewcommand{\shortauthors}{Kundu and Ayachitula et al.}

\maketitle

\begin{abstract}
  In this paper\footnote{This work was carried out when Ashish Kundu was working at IBM T J Watson Research Center, New York.}, we have studied how the integrity of the text of an ancient literature has been preserved for several centuries. Specifically, the Vedas is an ancient literature, which has its text remained preserved without any corruption for thousands of years. As we studied the system that protects the integrity of the text, pronunciation and semantics of the Vedas, we discovered a number of similarities it has with the current concept of blockchain technology. It is surprising that the notion of de-centralized trust and mathematical encodings have existed since thousands of years in order to protect this work of literature. We have presented our findings and analysis of the similarities. There are also certain technical mechanisms that The Vedic integrity system uses, which can be used to enhance the current digital blockchain platforms in terms of its security and robustness.

\end{abstract}

%%\begin{CCSXML}
%%<ccs2012>
%% <concept>
%%  <concept_id>10010520.10010553.10010562</concept_id>
%%  <concept_desc>Computer systems organization~Embedded systems</concept_desc>
%%  <concept_significance>500</concept_significance>
%% </concept>
%%</ccs2012>
%%\end{CCSXML}

%\ccsdesc[500]{Computer systems organization~Embedded systems}

%\keywords{Integrity, Blockchain, Consensus, The Vedas, Patha}



\section{Introduction}
Since the advent of bitcoin~\cite{bitcoin} and the underlying blockchain technology in 2008, the computing world have forked a foundational shift in the way certain computations and transaction processing are carried out. Several blockchain platforms – permission-ed and permission-less, have been developed; several applications as new cryptocurrencies have also been developed. Certain types of computing that involves untrusted/independent parties to collaborate and carry out operations or transactions based on a consensus require a computing framework that is supported by the blockchain technology.

Blockchain is often used for maintaining the integrity and provenance of historical records among multiple untrusted/independent peers. However, a pertinent question arises, when blockchain technology or even computing did not exist that how ancient texts authored maintained the integrity. One of the examples of such an ancient text are the Vedas~\cite{scharfe2002memorizing}.

We have observed that the integrity of the Vedas has been preserved for thousands of years by employing a process and technique that is de-centralized in nature. We think that such a process and technique is very similar to that of the blockchain technology~\cite{vo2018research}. Moreover, we also think that the system in place is akin to a hierarchical blockchain network – a network of networks.

\subsection{A Brief Summary of Blockchain}
Bitcoin~\cite{bitcoin} was proposed by Satoshi Nakamoto in 2008. He also introduced the concept of blockchain. Bitcoin is developed based on the concept of blockchain.  Bitcoin uses a public (permission-less) blockchain network that hosts multiple untrusted/independent peers for mining bitcoins (For further details on Bitcoin and blockchain, interested readers are referred to~\cite{bitcoin,bitcoinsok,vo2018research}).

In general, a blockchain network operates in a de-centralized manner. At a high-level, there are the following elements of blockchain system:

\begin{figure}[t]
  \centering
  \includegraphics[height=1.5in, width=3in]{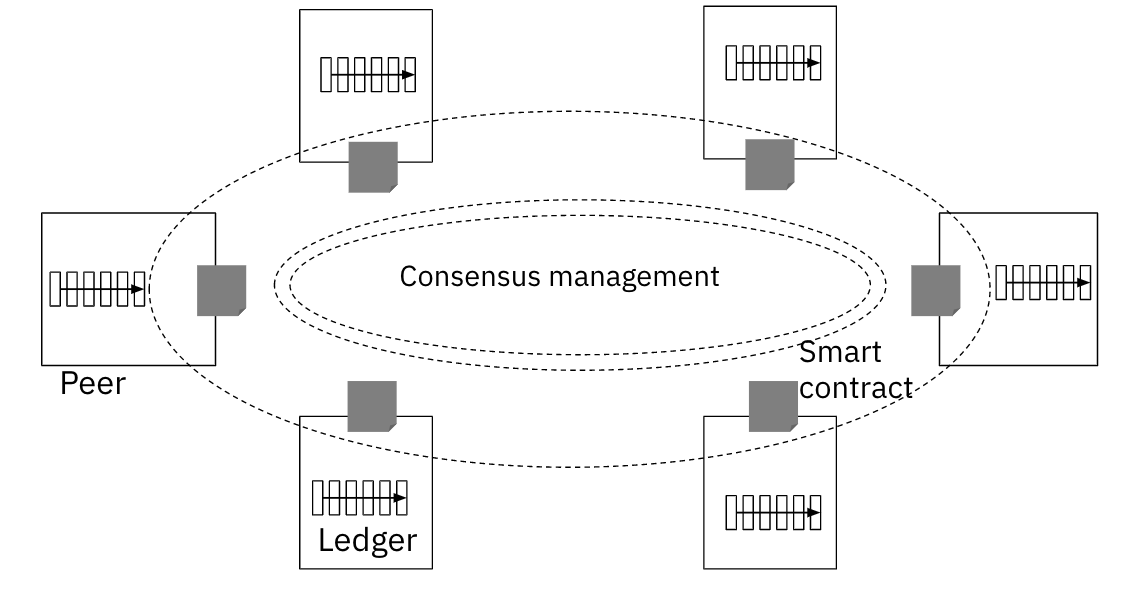}
  \caption{Blockchain Architecture.}
\end{figure}

\begin{itemize}
    \item Untrusted and independent peers
    \item Append-only, immutable ledger
\item Integrity mechanisms such as cryptographic hashing, digital signatures
\item Smart contract
\item Consensus management
\end{itemize}

Some of the several blockchain platforms and networks that are well-known today are:
\begin{itemize}
    \item Hyperledger Fabric
    \item Ethereum
    \item Hydrachain
\end{itemize}

\subsection{A Brief Summary of The Vedas}
The Vedas are known as the oldest literary texts. They are the scriptures of the Sanatana Dharma. The texts are composed in Sanskrit language. 
The Vedas have four Samhitas~\cite{TheVedas9}: Rigveda, Yajurveda, Samaveda, and Atharvaveda. There are about 89000 Padas and 72000 padas belong to the four Samhitas, which are also referred to as the Vedas in a collective manner.

The Vedas are known to be “Authorless” - Who authored it is not known. Vyasa Deva is regarded as the compiler of the Vedas~\cite{TheVedas2}. When was the Vedas authored is also not known and it is said that it is since time immemorial, but it is accepted that it has been known to exist for several thousands of years~\cite{TheVedas9}. Vedas are orally transmitted and are called as Shruti~\cite{TheVedas2}. The pronunciation of the Shlokas or Padas of the Vedas are foundational and are transmitted orally along with the text~\cite{wayne1986veda, UNESCO1, UNESCO2}.

\section{The Vedas and its Integrity}
In this section, we have presented the integrity protection system of The Vedas.

\subsection{Threat Model}
The threat model for The Vedas is described in this section. It is relatively easy for the text and pronunciations to get changed inadvertently or maliciously because  
\begin{enumerate}
    \item the Vedic text and pronunciations are transmitted orally since 1700 BCE (estimated)~\cite{TheVedas2}.
    \item there is no central authority that owns or administers the correctness, integrity and authenticity.
    \item the Vedas are so ancient that evolution of the society and human language forms may introduce or tamper with the pronunciations and the text.
\end{enumerate}

The potential threat actors could be anyone including the caretakers of The Vedas. However, the Vedas are known not to have incurred any textual corruption and tampering since thousands of years ~\cite{TheVedas9,scharfe2002memorizing}. It seems to be impossible, unless there is some intrinsic technology used by the Vedic experts and practitioners.

\subsection{Integrity Protection Mechanisms}
In this section we have described the integrity protection mechanism that The Vedas have employed for thousands of years if not more than that. 

Integrity of the Vedas is supported by the following elements:
\begin{itemize}
    \item patterns called as “Patha”
    \item one group of society preserve one pattern
    \item periodic recitation and consensus
\end{itemize}
`
\subsection{Mathematical Encodings: 'Patha'}
The Vedas use multiple systems of Vedic studies and recitals, and each system uses a specific pattern. Eleven different types of encodings are used and are called as 'patha' or pattern, where one or more padas are recited using each pattern~\cite{scharfe2002memorizing,UNESCO1,UNESCO2}.

Each pattern is called 'patha' or 'recital' represents a unique encoding of the terms in a verse.

\subsection{Patha}
Eleven types of pathas are used in encoding The Vedas' stanzas or Padas.
\begin{itemize}
    \item  Samhita, Pada, Krama, Ja\d{t}\={a}, M\={a}l\={a},\\ Sikh\={a}, Rekh\={a}, Dhvaja, Da\d{n}\d{d}a, Ratha, Ghana 
\end{itemize}

Each such encoding acts as a way to determine the integrity as well as errors if any during the transmission or recitals. A given patha has two dimensions – (1) textual/alphabetic (in Sanskrit~\cite{TheVedas2}) encoding, and (2) phonetic encoding. The special property here in the Vedas is that the encoding of a verse not only need to be accurate in the textual/alphabetic (in Sanskrit) representation of the patha but also the correctness in the pronunciation and its intonations. Within a given phrase - Sanskrit being an order free language it could be recited in any order without altering its meaning. 

Some authors have reported some other patterns or pathas to be used by The Vedas. We have used the ones mentioned in ~\cite{TheVedas2}.

If the original order of words in a pada is: 1/2/3/4/5/6/7/8/9

\emph{Ja\d{t}\={a} patha}: A recitation chain consists of forward-reverse-forward arrangement of words.

\begin{itemize}
    \item 1 2 2 1 1 2 / 2 3 3 2 2 3 / 3 4 4 3 3 4 / 4 5 5 4 4 5 /
\end{itemize}

\begin{figure}[t]
  \centering
  \includegraphics[height=0.3in, width=2in]{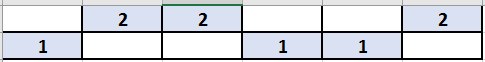}
  \caption{Ja\d{t}\={a} Patha.}
\end{figure}

\emph{Ghana patha}:  A recitation chain consists of one forward and reverse links of first two words followed by a forward-reverse-forward linking of the three words. 
\begin{itemize}
    \item 12/21/123/321/123/23/32/234/432/ \
    234/34/43/345/543/345/45/54/45/$\ldots$
\end{itemize}

\begin{figure}[t]
  \centering
  \includegraphics[height=0.5in, width=3in]{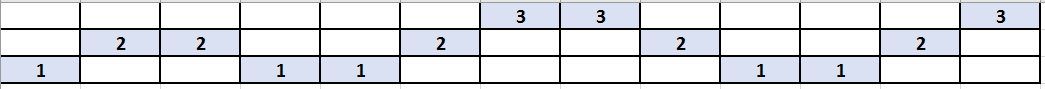}
  \caption{Ghana Patha.}
\end{figure}

\subsection{De-centralized Societal Groups}
The Vedic practitioners have eleven different groups. Each such group is assigned one specific patha and is responsible to transmit the Vedas in the specific patha. That specific group from generation to generation recites and transmits the Vedas in that pattern. For example, ghanapathis preserve the Ghana patha and recite and transmit the Vedas in the Ghana pattern.

If there any issues with the pronunciation or text of one or more verses, the individuals in a societal group can
\begin{enumerate}
\item determine that such an issue exists,
\item the pronunciation indicates one or more errors as one or more terms may not match the expected pronunciation.
\item the text pattern indicates one or more terms not matching the expected pattern.
\end{enumerate}

\subsection{Periodic Consensus}
Periodic conferences, meetings and gatherings among multiple members in the same group and/or across different groups are carried out. Recital of the Vedas in the one or more pathas may help validate the verses. Consensus mechanism that may employ a weighted consensus with perhaps more weight assigned to the vedic scholars, Gurus and Acharya's. The exact frequency, form, or process of such consensus is yet to be unearthed, but from our initial findings, we find that collective recitals often are carried out and that helps in automatic validation and verification of the correctness of the verses, the pronunciation and the semantics. We have been informed that there have been a prior instance where verses or pada's have been added to the Vedas. Perhaps such addition is carried out by a de-centralized consensus process because there is no central authority.
\begin{itemize}
    \item Periodic consensus helps validate, detect, (and correct, if possible)  
    \item errors or compromises due to
    \item invalid pronunciation and/or
    \item invalid verse
    \item invalid patha composition of the verse
    \item move to the next state by validating and permitting new additions (that occurs very infrequently – perhaps in hundreds or years, if not more).
\end{itemize}

\subsection{De-centralized Trust Model}
We studied the patterns and the social groups assigned to each pattern. It occurred to us that there is an inherent trust model built into the protection model of the Vedas.

Due to the lack of a central authority and separate social groups responsible for one patha, we infer  that the trust model supported by the Vedas for its transmission and integrity protection is de-centralized in nature.

\subsection{Integrity Protection and Analysis}
There are three types of checks that human beings perform on the Vedic padas during its recitals and chanting:
\begin{itemize}
    \item verification of text
    \item verification of pronunciation
    \item verification of semantics
\end{itemize}

Such a check is carried out by
\begin{itemize}
    \item one individual – centralized protection.
    \item by a group preserving one patha – one level of de-centralization.
    \item by a collection of some or all the groups and invoking consensus – higher than one level of de-centralization.
\end{itemize}

A tampering attack has to both satisfy the textual encoding, phonetic encoding and the semantics. By phonetic encoding, we mean that the Sanskrit pronunciation in that specific 'patha' needs to be accurate.

By a successful attack on the textual encoding, we mean that the patha cannot be used to determine if the text of the verse have been modified. Any change such as modifying the above pada to  1/2/3/4/5/6/7/8/9/10 by appending '10' after '9', there have to have several changes in the  rekhā recital.  

However, an error or a malicious tampering of a given pada may still be detected due to an incoherent pronunciation in the Sanskrit language. It is hard to pass this detection process. Assuming that a given attack is powerful and avoids detection at both textual and pronunciation detection phases. Semantics of a given verse, the terms and the semantics of the related other verses can be used to determine if there is such as corruption in a given verse. It is harder than the textual and pronunciation mechanisms to avoid detection at the semantics phase.

Moreover, if an attacker can bypass all the checks mentioned above for one individual, then the modified verse may not pass the group-level check for a given patha. If the modified verse passes the group-level check for a given patha, then it may not pass the consensus among some or all groups for the collective set of pathas. A successful attack has to pass through all these checks and balances, which is why perhaps the Vedas have continued to be protected textually, phonetically and semantically.

\subsection{Analogy and Connection with Digital Blockchain Technology}
Integrity protection of the Vedas is human-centric and society-centric, while today's blockchain technology is digital in nature. Integrity protection of the Vedas have been established since thousands of years even when there was no notion of digital computers, while the current digital blockchain technology is a recent development since 2008.

In this context, we summarize the integrity protection mechanisms for the Vedas:

\begin{itemize}
    \item de-centralized trust model
    \item multiple patterns called patha for encoding verses for error detection and integrity verification
    \item multiple groups of society – each group to preserve and recite for each pattern called patha
    \item integrity verification by verification of text encoding, pronunciation and semantics.
    \item consensus helps validate, detect, (and correct, if possible)    
\end{itemize}
 
 \begin{table*}[htb]
  \caption{Analogy between The Vedic Blockchain Model and today's Digital Blockchain}
  \label{tab:commands}
  \centering
  \begin{tabular}{|l|l|}
  \hline
    Human Blockchain of The Vedas & Digital Blockchain\\
    \hline
    De-centralized trust model & De-centralized trust model\\
    \hline
    Integrity supported by patterns called as “Patha” & Hashing\\
    \hline
    Eleven different types of patterns: & \\
    Samhita, Pada, Krama, Ja\d{t}\={a}, M\={a}l\={a}, & \\ Sikh\={a}, Rekh\={a}, Dhvaja, Da\d{n}\d{d}a, Ratha, Ghana & Multiple Hashing Schemes\\
    \hline
    One group of society preserve one pattern & Peers, Miners\\
    \hline
    Consensus among the individuals & Consensus management across\\ in a group, among groups & peers, miners, lineage, provenance, immutability\\
    \hline
  \end{tabular}
\end{table*}

\section{Blockchain based on Multi-hashing schemes}
Digital blockchain supports only one hashing technique or a digital signature technique. However, the Vedas employ multiple encoding schemes (each is a patha). That is more robust than using one patha because if one patha is compromised, then a tampering attack is successful. 

Today's digital blockchain should use multiple hashing schemes or digital signature schemes in order to support better security for the hashchain and the ledger, and thus for the overall blockchain network.
If the blockchain supports digital signature scheme, it may support multiple separate signature schemes.

%\section{Discussions}
%\input{discussions}

\section{Conclusions \& Future Work}
In this paper, we have described an interesting finding. We have discovered that the ancient text of the Vedas has been protected from errors and integrity attacks on its text, pronunciation and semantics by using a de-centralized trust model and a blockchain-like system. It is surprising that since thousands of years a blockchain-like system is in place. However, some of these findings and analogy here requires further research and analysis.
%\begin{acks}
%\end{acks}

\bibliographystyle{abbrv}
%\bibliography{sample-base}
\bibliography{TheVedas-Blockchain}

%\appendix

\end{document}